\documentclass[twocolumn,preprintnumbers,amsmath,amssymb]{revtex4-2}
\usepackage{graphicx}
\usepackage[normalem]{ulem}
\usepackage[svgnames]{xcolor}
\usepackage{bm}
\usepackage{multirow}
\usepackage{float}
\usepackage{titlesec}
\usepackage[columnwise]{lineno}
\usepackage{float}
\usepackage{titlesec}
\usepackage{gensymb}
\usepackage{textcomp}
\begin{document}
\title{On the origin and the amplitude of T-square resistivity in Fermi liquids}
\author{Kamran Behnia}
\affiliation{LPEM (CNRS-Sorbonne University), ESPCI Paris, PSL University, 75005 Paris, France}
\date{\today}
\begin{abstract}
In 1937, Baber, Landau and Pomeranchuk postulated that collisions between electrons generates a contribution to the electric resistivity of metals with a distinct T$^2$ temperature dependence. The amplitude of this term is small in common metals, but dominant in metals hosting either heavy carriers or a low concentration of them. The link between the temperature dependence and the size of the scattering phase space is straightforward, but not the microscopic source of dissipation. To explain how electron-electron collisions lead to momentum leak, Umklapp events or multiple electron reservoirs have been invoked. This interpretation is challenged by a number of experimental observations: the persistence of T-square resistivity in dilute metals (in which the two mechanisms are irrelevant), the successful  extension of Kadowaki-Woods scaling to dilute metals, and the observation of a size-dependent T-square thermal resistivity ($T/\kappa$) and its Wiedemann-Franz (WF) correlation with T-square electrical resistivity. This paper argues that much insight is provided by the case of normal liquid $^3$He where the T-square temperature dependence of energy and momentum diffusivity is driven by fermion-fermion collisions. The amplitude of T-square resistivity in $^3$He and in metals share a common scaling. Thus, the ubiquitous T-square electrical resistivity  ultimately stems from the Fermi-liquid temperature dependence of momentum diffusivity.
\end{abstract}
\maketitle
The contribution of electron-electron scattering to the temperature dependence of electrical conductivity in conducting solids has a long history, which begins in the late 1930s with its  theoretical identification by Baber \cite{Baber1937} and (contemporaneously and independently) by Landau and Pomeranchuk \cite{Baber1937, Landau1937}. The concept began to attract significant interest several decades later \cite{Rice1968,Kaveh1984} with  the accumulation of experimental data. In common metals, this effect is not easy to observe and remains small compared with the contribution due to electron-phonon scattering. The discovery of metals hosting strongly-correlated electrons changed the experimental landscape in 1980s. In heavy-fermion systems \cite{Fisk1986} electron-electron scattering dominates the temperature dependence of resistivity at cryogenic temperatures. Kadowaki and Woods \cite{KADOWAKI} (following an idea first put forward by Rice about transition metals \cite{Rice1968}) observed that the prefactor of T-square resistivity scales with the square of the electronic specific heat. This scaling, based on the argument that both quantities are amplified by enhancement in the density of states proved influential during decades of experimental research on heavy fermion metals (for an early example see \cite{Amato1987}). It was also used as a road-map to track the emergence of what was dubbed `Non-Fermi-liquid behavior' in strongly-correlated-electron systems \cite{Stewart2001}. The microscopic origin of dissipation did not attract much debate. It was assumed that Umklapp events allowed by the large Fermi surface of the metallic solids of interest are what drive the T-square resistivity \cite{Maebashi}.  

Recently, a modified version of the  Rice or Kadowaki-Woods scaling was proposed to explain the amplitude of the T-square prefactor in dilute metals, which are zero-temperature conductors hosting a very low concentration of mobile electrons \cite{lin2015,collignon2019,Wang2020}. These are either stoichiometric semi-metals (such as bismuth) or semiconductors sufficiently doped to be on the metallic side of the metal-insulator transition (such as oxygen-reduced strontium titanate). In their case, the relevant scaling is between the prefactor of T-square resistivity and the inverse of the square of the Fermi energy (and not the specific heat which is contaminated by the unusually low carrier density). 

Two new experimental developments have accompanied the renewal of interest for this theme (see ref.\cite{Pal2012} for a review of possible origins of T-square resistivity prior to them). The first is the observation that T-square resistivity persists even when the microscopic origin of the dissipation mechanism cannot be identified \cite{lin2015,Wang2020}. The second is the observation of T-square \textit{thermal} resistivity in semi-metals close to the ballistic regime \cite{jaoui2018,Jaoui2021}.  They raise a number of  questions: Is the T-square resistivity in dilute metallic strontium titanate \cite{Tokura1993,Marel2011,lin2015,Mikheev2016} due to electron-electron scattering ? If yes, given the absence of Umklapp events and multiple pockets, what causes it? If not, what kind of electron-phonon scattering mechanism can generate a T-square resistivity \cite{Kumar2021}? A similar question was raised long ago in the case of bismuth. The  observed T-square resistivity to was attributed to scattering between electrons residing in different pockets\cite{Hartman1969}. But this was subsequently contested by other authors \cite{Uher1977,kukkonen}. It is raised again by the observation of T-square resistivity in another dilute metallic solid, namely Bi$_2$O$_2$Se \cite{Wang2020}, implying that SrTiO$_3$ \cite{lin2015} is not a lonely anomaly. More fundamentally and beyond these specific cases,  What is the reason behind the success of  scaling \`a la Kadowaki-Woods across a wide variety of families of materials (provided that one takes in to account the carrier density)? How does this fit with the assumption that the microscopic mechanism generating momentum decay from electron-electron collisions varies from one case to the other? 

The aim of the present paper is to argue in favor of a route from electron-electron scattering to T-square resistivity in metals starting from our understanding of thermal transport in liquid $^3$He \cite{Abrikosov_1959,Nozieres,Brooker1968,Sykes1970,wolfle1979}. It has been known for decades that, well below the degeneracy temperature of $\sim 1.5 K$, collisions between fermions in liquid $^3$He generate a thermal conductivity which decreases with the inverse of temperature \cite{Wheatley,abel1967,greywall1984,calkoen1986}. This is strictly equivalent to the T-square thermal resistivity of electrons, as defined by metal physicists. This picture has no need for Umklapp events, or a bottleneck provided by multiple fermionic reservoirs. Its relevance to electrons in metals was experimentally demonstrated in the case of semi-metallic antimony \cite{Jaoui2021}. Its potentially broader relevance would solve a number of pending puzzles in other specific materials.  

I will begin by a short historical account of experimental and theoretical investigations of T-square electrical resistivity. Then, I will recall (much less frequent) studies detecting the thermal counterpart of T-square resistivity.  The next section is devoted to normal liquid $^3$He. Greywell's extended study of the evolution of specific heat \cite{greywall1983} and thermal conductivity \cite{greywall1983} with pressure is reviewed.  It remains the most detailed experimental verification of the theory of momentum and energy diffusivity in Fermi liquids. By comparing electrons in metals (in presence of lattice and its disorder) and fermions in $^3$He (in absence of both) a common scaling emerges. This leaves little doubt regarding the main driver of T-square thermal resistivity. The neglected role of viscosity of the electron liquid \cite{kishore,giuliani_vignale_2005} in the emergence of  T-square electrical resistivity is highlighted by this analogy.

\section{Ubiquity and scalability of T-square resistivity in Fermi liquids}
\begin{figure*}
\centering
\includegraphics[width=15 cm]{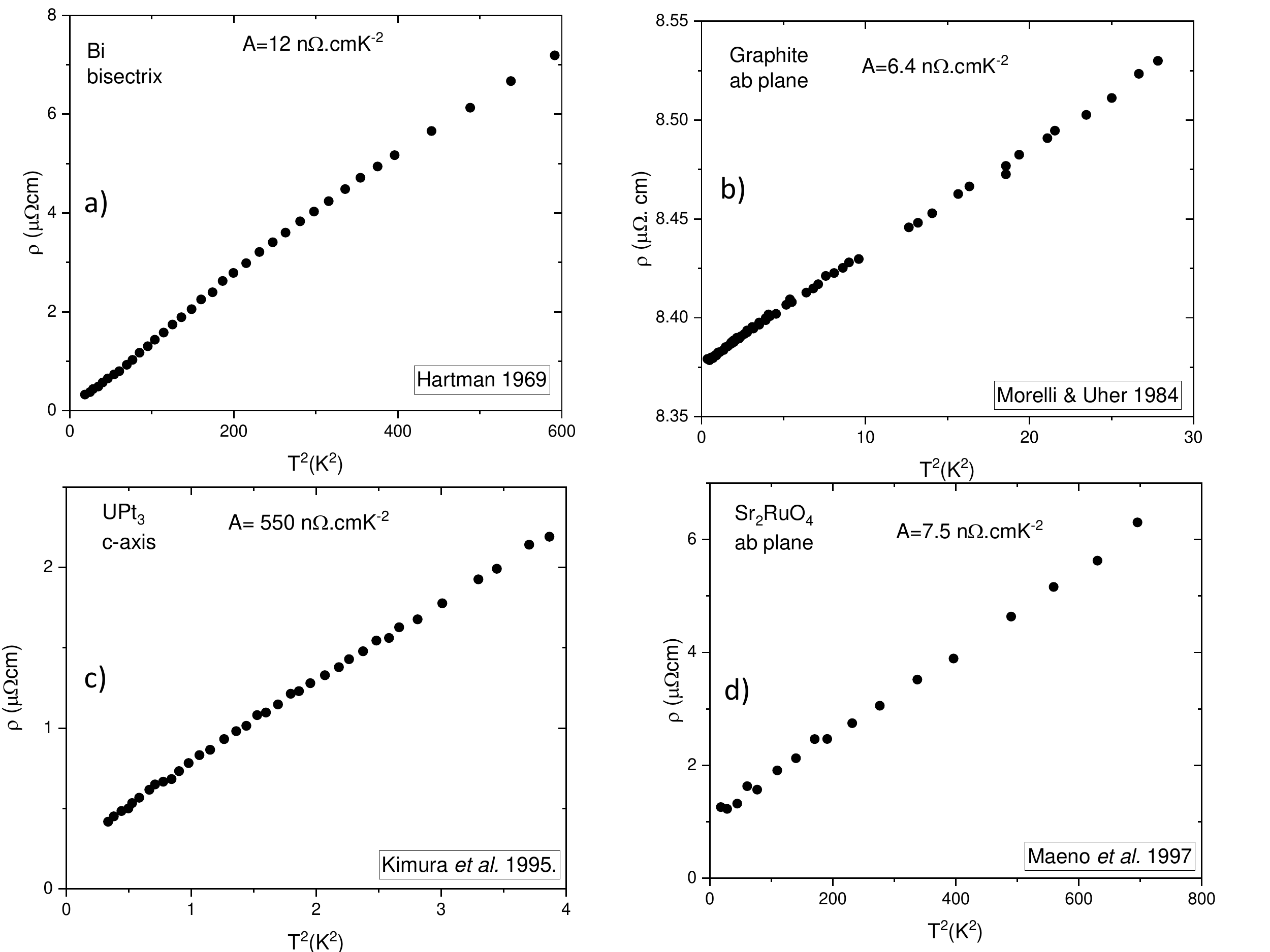}
\caption{\textbf{Examples of T-square resistivity:} Resistivity as function of $T^2$ in bismuth \cite{Hartman1969}(a), in graphite \cite{Uher1977} (b), in UPt$_3$ \cite{Kimura} (c) and in Sr$_2$RuO$_4$ \cite{Maeno1997} (d).}
\label{fig:examp}
\end{figure*}

Fig.\ref{fig:examp} shows the low-temperature resistivity of four different metals. Bismuth and graphite are elemental semi-metals. Sr$_2$RuO$_4$ and UPt$_3$ are correlated metals each with an unconventional superconducting ground state. Low-temperature resistivity of these four solids quadratically increases with temperature, following :
\begin{equation}
 \rho=\rho_0+AT^2 
 \label{rho}
\end{equation}

The first term on the right side is residual resistivity, $\rho_0$, attributed to impurity scattering. The second term, $A$ (in ohm.m.K$^{-2}$), is attributed to electron-electron scattering. It will be the focus of this paper. The low-temperature resistivity of these four solids lacks any trace of $T^5$ resistivity caused by electron-phonon scattering in the Bloch-Gr\"uneisen picture \cite{ziman1972}. This is because $A$ is large enough to impede its observation. In contrast, $A$ is orders of magnitude smaller in  noble or alkali metals and temperature-dependent resistivity is dominantly  $T^5$. 
\begin{figure*}
\centering
\includegraphics[width=15 cm]{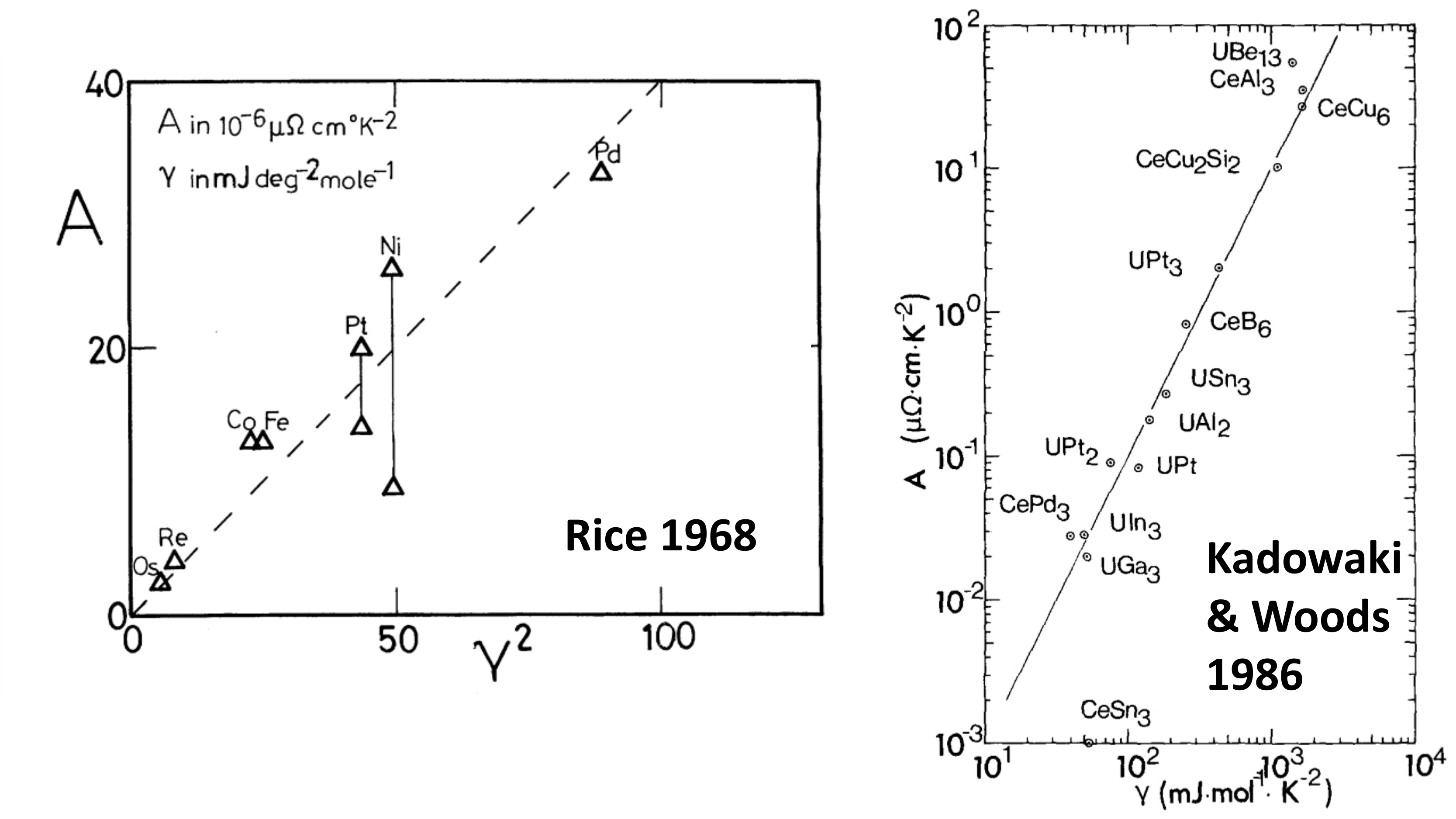}
\caption{\textbf{$A$ \textit{vs.} $\gamma^2$ scaling:} Left:The amplitude of $A$ in several transition metals it scales with $\gamma^2$. Rice's 1968 figure \cite{Rice1968} [reproduced with  permission granted by APS].  Right: Kadowaki and Woods \cite{KADOWAKI} found a similar scaling in the case of heavy-fermion metals [reproduced by with permission granted by Elsevier]. Dilute  metals are absent in both plots.}
\label{fig:rice}
\end{figure*}

What sets the amplitude of $A$ in a given metal? Does it correlate with $\rho_0$? Numerous experiments point to a negative answer to the latter question and indicate that $A$ is an intrinsic property of a Fermi liquid. In many  heavy-fermion metals, the quality of crystals have improved over the years, pulling down $\rho_0$ by  more than one order of magnitude, yet  $A$ is similar in clean and dirty crystals. In URu$_2$Si$_2$, for example,in 1986, Palstra \textit{el al.} \cite{Palstra1986} found $\rho_0= 33 \mu \Omega.cm $  and $A=0.1 \mu \Omega.cmK^{-2}$. A quarter of century later, working on much cleaner crystals, Matsuda \textit{et al.} \cite{Matsuda2011} reported  $\rho_0= 1.05 \mu \Omega.cm $  and $A=0.099\mu \Omega.cmK^{-2}$. The same is true for UPt$_3$, where the magnitude of $A$ remains the same in crystals in which $\rho_0$ differe by one order of magnitude \cite{Joynt2002}. One can go also along the opposite direction by introducing disorder through electronic irradiation, as in the case of metallic strontium titanate, where a twofold enhancement in $\rho_0$ leaves $A$ unchanged \cite{Lin2015b}.

The intrinsic nature of $A$ is backed by the success of scaling approaches, which was initiated  by Rice in 1968 \cite{Rice1968}. He demonstrated that $A$ scales with the square of the electronic T-linear specific heat $\gamma$ (whose unit is J.K$^{-1}$.mo$l^{-2}$) in seven elemental transition metals. 18 years later, Kadowaki and Woods (KW) \cite{KADOWAKI} applied this scaling to the newly-discovered Heavy-fermion metals and found that a similar scaling works there too. They  also noticed a difference in the amplitude of the ratio. In heavy fermions, it is  ($\frac{A}{\gamma^2}=10^{-7}\Omega$.m($\frac{mol.K}{J})^2$). This is more than one order of magnitude larger than what Rice had found in transition metals ($\frac{A}{\gamma^2}=4 \times 10^{-9}\Omega$.m($\frac{mol.K}{J})^2$). Theses historical plots are reproduced in Fig. \ref{fig:rice}.  

This difference between strongly-correlated and weakly-correlated systems became fuzzier as the data became more abundant. In 2003, Tsuiji \textit{et al.} \cite{tsujii2003} showed that most Yb-based  heavy fermions are closer to the Rice scaling line than to the KW scaling line. Nevertheless, this scaling approach is impressively effective over eight to nine orders of magnitude. It implied that one could predict the rough amplitude of one measurable quantity (namely $A$), thanks to knowing the amplitude of another measurable quantity (namely $\gamma$). 
\begin{figure*}
\centering
\includegraphics[width=13.5 cm]{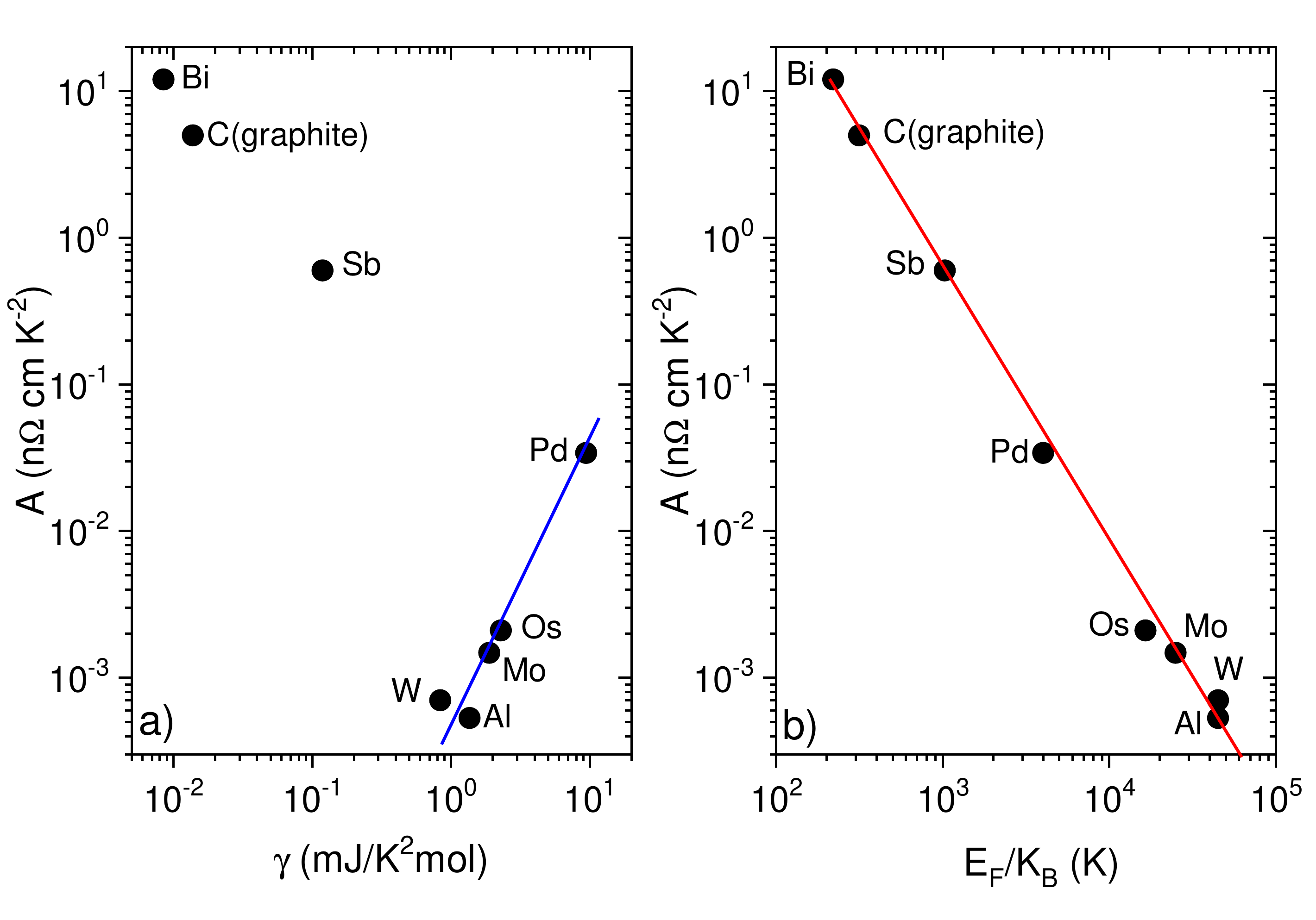}
\caption{\textbf{The proper scaling:} a) $A$ \textit{vs.} $\gamma$ in several metallic elements including low-density semi-metals. The blue solid line represents $A \propto \gamma^2$. b) $A$ of the same metallic elements \textit{vs.}  the Fermi energy, $E_F$, in kelvin. The red solid line represents $A \propto E_F^{-2}$. }
\label{fig:rice-bis}
\end{figure*}

The Pauli exclusion principle is the ultimate reason behind both the temperature dependence of electronic specific heat and electron-electron contribution to the resistivity. Electrons giving rise to both properties are those confined to a thermal window near the Fermi energy. This leads to the following expression for the electron specific heat (per volume):

\begin{equation}
\frac{C_e}{T}= \gamma = \frac{\pi^2}{2}\frac{nk_B^2}{E_F} 
\label{gamma}
\end{equation}

Here $n$ is the carrier density. Now, the phase space for collisions between two electrons is proportional to $(\frac{k_BT}{E_F})^2$. Since $A$ is expressed in $\Omega$m K$^{-2}$, it can be written as \cite{lin2015}:  

\begin{equation}
A= \frac{\hbar}{e^2}(\frac{k_B}{E_F})^2 \ell_{quad}
\label{A}
\end{equation}

Here, $\hbar$ and $e$ are the Planck constant (divided by $2\pi$) and the fundamental charge. The material-dependent parameter $\ell_{quad}$ has the dimension of a length and condenses all the microscopic details of scattering.  

According to these equations, $\gamma \propto$ E$_F^{-1}$ and  A $\propto$ E$_F^{-2}$. As a consequence,  the A$\propto \gamma^2$, provided that $\ell_{quad}$ does not drastically vary from one metal to another. At the same time, the scaling should collapse when the comparison is made between metals with  widely different carrier densities. As seen in Fig.\ref{fig:rice-bis}a, this is indeed the case. As soon as one adds bismuth (Bi), graphite (C) and antimony (Sb), to Rice's plot, the scaling does not work. In these semi-metals, a single electron is shared by thousands of atoms and the magnitude of $\gamma$ is pulled down by the low carrier density. As a consequence, $A$ decreases with $\gamma$.

In dilute metallic strontium titanate, where carrier density can be tuned by many orders of magnitude, the simple KW scaling between $A$ and $\gamma$ does not hold either \cite{McCalla}. The electronic specific heat with doping depends both on the effective mass, which increases by a factor of $\approx$ 2, but much more on the carrier density, which changes by several orders of magnitude. The Sommerfeld coefficient, $\gamma$ is compatible with the experimental quantum oscillations \cite{Lin2014b} and calculated band structure \cite{Marel2011}. The magnitude of $A$ matches what was calculated by Hussey \cite{Hussey2005}. Nevertheless, $A$ decreases with $\gamma$ as it does when comparing Bi, graphite and Sb in Fig. \ref{fig:rice-bis}a.

To include dilute metals in this scaling scheme, it is necessary to replace $\gamma^2$ by $E_F^2$ as the horizontal axis \cite{lin2015}. The Fermi energy is a quantity accessible to the experimentalist. It can be quantified by measuring the Fermi radius and the effective mass of electrons, or by a combined knowledge of $\gamma$ and $n$. It is true that a multi-band metal has several Fermi energies. In semi-metals, the Fermi energy of electrons and holes differ, because of the mismatch in their masses. However, since we are dealing with orders of magnitude, the average Fermi energy can do the job.

Fig.\ref{fig:rice-bis}b, is a plot of $A$ \textit{vs.} $E_F^2$ for the same elements plotted in Fig.\ref{fig:rice-bis}a. They are listed together with their relevant specific properties in table \ref{Tab1}. The scaling operates across more than five orders of magnitude of $A$. This scaling was first put forward in ref.\cite{lin2015} and was extended in ref. \cite{collignon2019} and ref.\cite{Wang2020}. 

The simple KW  scaling (i.e $A/\gamma^2$) for correlated metals  fails also for dilute metals. A revealing  case is CeNiSn, a low-density Kondo semi-metal \cite{Nakamoto1995}. The prefactor of its T-square resistivity is $A=2.2 \mu \Omega$cmK$^{-2}$,  two orders of magnitude larger than what is expected according to its electronic specific heat, which is  $\gamma = 40 $mJ.mol$^{-1}$K$^{-2}$ \cite{Izawa1999}  combined with the KW  $\frac{A}{\gamma^2}$ universal ratio. However, this discrepancy evaporates if one extracts the Fermi energy from the slope of the Seebeck coefficient \cite{Behnia_2004}, as noticed by Kuitra \textit{et al.} \cite{Kurita2019}. The latter is indeed adequately large in CeNiSn \cite{HIESS}.  

Hussey \cite{Hussey2005} was the first to put his finger on the inadequacy of the simple KW scaling. He was partly motivated by the case of Na$_{0.7}$CoO$_2$, in which an unexpectedly large $A/\gamma^2$ was resolved \cite{Li2004}.  He proposed a modified version of the KW scaling in which  $A$ scales with the electronic specific heat per volume and not per mole. However, finding a general expression for $A$ remains a challenge, since it is  transport coefficient and cannot be derived out without specific and contestable assumptions about scattering details. An alternative amendment to the KW scaling \cite{Jacko2009} consisted in abandoning the idea of scaling $A$ with a quantity directly accessible to the experiment. 

Empirically, however, Eq. \ref{A} gives a reasonable account of $A$ in numerous metals over eight orders of magnitude \cite{Wang2020}. The list includes numerous newly discovered semi-metals, such as WTe$_2$ \cite{zhu2015}, which have attracted attention for other reasons. The material-dependent  $\ell_{quad}$ resides in the range of 1-40 nm\cite{Wang2020}. Each family of metal clusters along specific values of $\ell_{quad}$. In elemental metals of Fig.\ref{fig:rice-bis}, $\ell_{quad} \approx 2$nm. In layered metals, such as Sr$_2$RuO$_4$ \cite{Maeno1997} or  La$_{1.7}$Sr$_{0.3}$Cu0$_4$\cite{Nakamae2003}, the out-of-plane resistivity is orders of magnitude larger the in-plane resistivity. As a consequence, $A$ and $\ell_{quad}$ are also anisotropic. In these cases, our focus will be the in-plane transport.

\begin{table*}
\centering
\begin{tabular}{lccccc}
\hline
Element & Carrier density (m$^3$) & $\gamma$ (mJ.mol$^{-1}$.K$^{-2})$ & $E_F$ (K)& $A$ (n$\Omega$.cmK$^{-2}$) & ref. \\
\hline
Bi & 6$\times 10^{23}$ & 0.0085& 220 & 12& \cite{Hartman1969,issi1979,collan1970,liu1995} \\
C(graphite) & 6$\times 10^{24}$ & 0.0138& 315 & 5& \cite{Hoeven1963,Uher1977,Brandt} \\
Sb & 1.1$\times 10^{26}$ & 0.119& 1030 & 0.6& \cite{collan1970,issi1979,Jaoui2021,liu1995} \\
Mo & 2.5$\times 10^{28}$ & 1.9& $2.5\times 10^4$ & $1.5\times 10^{-3}$& \cite{Bryant,Desai1984} \\
W & 1.4$\times 10^{28}$ & 0.84& $4.5\times 10^4$ & $7\times 10^{-4}$& \cite{White1984,GIRVAN19681485,Desai1984} \\
Pd & 5$\times 10^{28}$ & 9.43& $4\times 10^3$ & 0.034& \cite{Rice1968,Wilding_1967} \\
Al & 6$\times 10^{28}$ & 1.37& $4.5\times 10^4$ & $5.3\times 10^4$& \cite{Parkinson_1958,AMUNDSEN1967718,Garland_1978} \\
\hline
\end{tabular}
\caption{The experimentally resolved $A$ in a number of elements together with a number of relevant physical properties. In  semi-metals (Bi, graphite, Sb, Mo and W), the carrier density is the sum of the density of electrons and holes and the Fermi energy is the average of the Fermi energy of the holes and electrons.}
\label{Tab1}
\end{table*}

This brings us to the physical significance of this phenomenological length scale. Mott \cite{Mott1990} argued that $A$ is proportional to the collision cross section of the two electrons, $\sigma_{cs}$. Therefore:
\begin{equation}
\ell_{quad}\propto k_F \sigma_{cs}
\label{ell2}
\end{equation}

The Fermi wavevector, $k_F$, and the collision cross-section $\sigma_{cs}$ are expected to evolve in opposite directions with the evolution of carrier density. In a dilute metal, the Fermi wave-vector is short and the collision cross section (in real space) is large. The opposite is true in a dense metal. We note that $\ell_{quad}$ in Bi and Al are roughly  similar. We will come back to this question in section 5.

A fundamental question is the following: Why should electron-electron collision decay the charge flow ? Answers to this question fall in to two categories. 

First: If the wave-vector of one of the two electrons after the collision is larger than the width of the Brillouin zone, then a finite amount of momentum (equal to $\hbar G$, where $G$ is a unit vector of the reciprocal lattice) will be lost. This Umklapp option was what Landau and Pomeranchuk had it in mind in postulating T-square resistivity \cite{Landau1937}. Umklapp events require a sufficiently large Fermi surface, which is the case of all metals, put under scrutiny by Rice \cite{Rice1968} and by Kadowaki and Woods\cite{KADOWAKI}. In the case of heavy-fermion metals the role of Umklapp events in generating $T^2$-ressitivity was explicitly recognized by Maebashi and Fukuyama \cite{Maebashi}.

Second: If there are two distinct electron reservoirs, then momentum exchange between these two reservoirs may constitute a bottleneck for momentum loss. This was the scenario explicitly put forward by Baber \cite{Baber1937} and was invoked by Hartman \cite{Hartman1969} in the case of bismuth in which the Fermi pockets are too small to allow Umklapp events. 

However, none of these two options work for dilute metallic strontium titanate \cite{lin2015}, which motivated a search for alternative mechanisms of T-square resistivity \cite{Kumar2021} generated by electron-phonon and not electron-electron scattering. It is true that strontium titanate is a non-trivial solid with peculiar phonon modes \cite{Yamada1969b}. However, it is not the only case of T-square resistivity in absence of Umklapp or multiple pockets. Bi$_2$O$_2$Se\cite{Wang2020} is yet another. Moreover, specifically tailored scenarios fail to explain the success of the scaling approach across many orders of magnitude irrespective of the presence or absence of Umklapp events.

In search of an alternative solution, let us turn our attention to heat transport.  
\section{The Wiedemann-Franz law and the T-square thermal resistivity}
In the zero temperature limit, thermal conductivity, like specific heat and in contrast to electrical conductivity, vanishes. Disorder, or finite size lead to a finite residual electrical resistivity, $\rho_0$ and a residual T-linear thermal conductivity, $\frac{\kappa_0}{T}$. Their ratio obeys the Wiedeamnn-Franz law :  

\begin{equation}
\rho_0\frac{\kappa_0}{T}=\frac{\pi^2}{3}\frac{k_B^2}{e^2}
\label{WF}
\end{equation}

$L_0=\frac{\pi^2}{3}\frac{k_B^2}{e^2}$ is known as the Sommerfeld value. The validity of this equality has been experimentally verified in all bulk metals hitherto investigated. 

\begin{figure*}
\centering
\includegraphics[width=15cm]{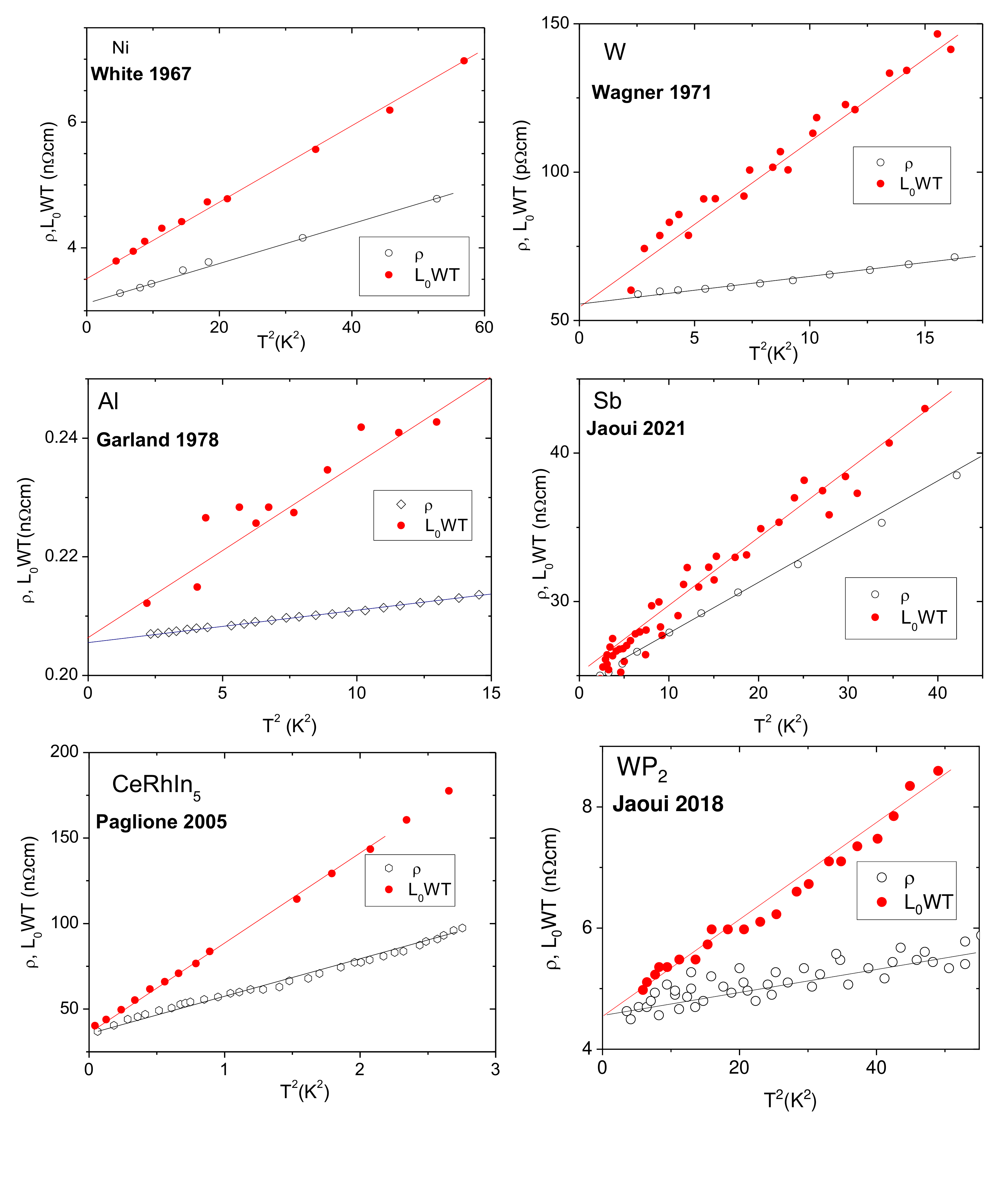}
\caption{\textbf{T-square thermal resistivity in several metals:} Low-temperature electrical resistivity, $\rho$ and thermal resistivity normalized by the Sommerfeld value, $L_0WT$ (in order to be expressed in the same units), for Ni \cite{White1967}, Al \cite{Garland_1978}, W \cite{wagner1971}, Sb \cite{jaoui2018}, CeRhIn$_5$ \cite{paglione2005} and WP$_2$ \cite{Jaoui2021}. In all cases, the two data sets join as the temperature tends to zero. In all cases, the slope of the thermal resistivity is higher. }
\label{fig:WT}
\end{figure*}

The fact that the phase space for electron-electron scattering grows quadratically with temperature has a signature in thermal transport. It leads to a thermal conductivity to the inverse of the temperature. In order to make a parallel with electrical resistivity, it is convenient to define thermal resistivity as $WT=\frac{T}{\kappa}$ and then, one has:
\begin{equation}
 WT=WT_0+BT^2 
\label{WT}
\end{equation}
The T-square prefactor, $B$, is the thermal counterpart of $A$ in Eq. \ref{rho}. Several experiments have quantified $B$ in a number of metals \cite{White1967,wagner1971,Garland_1978,Behnia1991,lussier1994,uher2004,paglione2005,paglione2006,seyfarth2008,jaoui2018,Jaoui2021}. These studies remain much less numerous than those devoted to charge transport. A selection of their findings is shown in Fig. \ref{fig:WT}. 

These experiments find that $L_0WT_0=\rho_0$, but $BL_0 > A$. In other words, The WF law is valid in the zero -temperature limit, but not at finite temperature. 

Historically, this finite-temperature deviation from the WF correlation has been understood as a specific manifestation of the different responses of momentum and energy flow to inelastic collisions \cite{ziman1972}. The starting point is a distinction between `vertical' and 'horizontal' collision events, A `horizontal event', in which the loss of energy is concomitant with a loss of momentum, would affect both electric and thermal conductivity. However, one can conceive `vertical' events, where a `hot' electron loses the energy it had to occupy a state slightly above the Fermi energy, without significant modification of its momentum vector. Such events, which require finite temperature, have little effect on electric resistivity, but drastically reduce  the thermal resistivity \cite{ziman1972}.

The distinction between `vertical' and 'horizontal' is also relevant to electron-phonon scattering. The resistivity of noble metals is dominated by electron-phonon scattering and follows a $T^5$ temperature dependence. Their  Lorenz number becomes significantly lower than the Sommerfeld value at intermediate temperature range \cite{White1960,yao2017}. The WF law is eventually recovered above the Debye temperature, when scattering  becomes elastic. 

There is an important difference between being scattered by phonons and being scattered by other electrons, however. In the case of e-ph scattering, the temperature dependence of the phase space differs for energy and momentum diffusion. Thermal resistivity follows $T^3$, the temperature dependence of the phonon population. Electrical resistivity, on the other hand, follows $T^5$. This is because the fraction of the phonons capable of inducing 'horizontal' events shrinks by an additional factor of $T^2$. The $T^5$  e-ph electrical resistivity is the low-temperature asymptotic behavior expected in the Bloch-Gr\"unesisen model \cite{ziman1972}, successfully applied to the experimental data in numerous elemental metals.  The $T^3$ e-ph thermal resistivity ($WT \propto T^3$, i.e. $W \propto T^2$) have been experimentally resolved in  several elemental metals \cite{Mendelssohn_1952,Rosenberg1955} as well as in the semi-metallic WP$_2$ \cite{jaoui2018}. 

In contrast, the phase space for e-e scattering follows $T^2$ for both thermal and electric resistivity. The differentiation between 'vertical' and 'horizontal' scattering would only lead to a smaller prefactor for electrical resistivity.  Herring argued that this difference is bounded to a value less than 2, that is $BL_0/A <2$. Li and Maslov, on the other hand, argued that in a compensated semi-metal $BL_0/A$ can become arbitrarily small, if the screening length is long enough compared to the Fermi wavelength \cite{Li-Maslov}.  These theoretical proposals have not been confirmed by the experimental data. The bound proposed by Herring is violated in several cases, such as W ($BL_0/A \approx 6$) \cite{wagner1971} and WP$_2$ ($BL_0/A \approx 5$) \cite{jaoui2018}. Moreover, the largest reported deviation from the WF law occurs in Al ($BL_0/A \approx 10$) \cite{Garland_1978}, which is \textit{not} a compensated semi-metal.

T-square thermal resistivity, in contrast with T-square electric resistivity,  does not require Umklapp events or multiple reservoirs. Normal e-e scattering inside a Fermi pocket can generate T-square resistivity. This is another dichotomy, distinct from horizontal and vertical events and capable of playing a role in the observed downward deviation from the Wiedemann-Franz law. Let us consider the case of thermal transport in a liquid of neutral fermions without the annoying presence of a lattice. 

\section{Energy and momentum diffusivity in $^3$He}
Soon after the conception of Landau's Fermi liquid theory \cite{Landau1957}, Abrisokov and Khalatnikov wrote a paper on the kinetic properties of a Fermi liquid with the case of $^3$He in mind \cite{Abrikosov_1959}. In this paper, they derived expressions for the temperature dependence of viscosity and thermal conductivity.  

Kinematic viscosity is the diffusion constant for momentum and thermal diffusivity is the diffusion constant for energy. In the simple kinetic theory for classical gases, there is a simple expression for diffusivity :

\begin{equation}
 D= \frac{1}{3}\tau v^2_m
\end{equation}

Here, $v_m$ is the mean molecular velocity and $\tau$, the mean scattering time. In a degenerate fermionic fluid,  $v_m$ becomes the Fermi velocity and $\tau$ is the inverse of the fermion-fermion scattering rate, which follows $T^{-2}$. Thus, both the viscosity and the thermal diffusivity are expected to decrease quadratically with temperature. Now, thermal conductivity is proportional to the product of (energy) diffusivity and the specific heat and the latter is T-linear. Therefore, one expects a thermal conductivity following $T^{-1}$.  

The temperature dependence obtained by such simplistic arguments are confirmed by the outcome of the elaborate calculations performed by Abriokosov and Khatalnikov \cite{Abrikosov_1959}. They  quantified the expected prefactors for the  $T^{-2}$ viscosity and the $T^{-1}$ thermal conductivity. Brooker and Sykes \cite{Brooker1968,Sykes1970} revisited the subject a decade later and found that the previous result for theoretical thermal conductivity was too large by a factor of 2. Moreover, they performed the first comparison between theory and experiment and found \cite{Brooker1968} their theoretical $\kappa T$ to be still 1.5 times larger than the experimentally-measured $\kappa T$ reported by Wheatley and his collaborators \cite{abel1967,Wheatley}. A year after, Dy and Pethick \cite{Dy1969} carried out a new set of calculations (assuming negligible Landau parameters when $\ell >2$) and found a closer agreement with the experimental data.

An extensive study of thermal conductivity in $^3$He was published by Greywall in 1984 \cite{greywall1984}, a year after his study of specific heat \cite{greywall1983}. They document in detail how entropy transport evolves with the strength of correlations in this emblematic Fermi liquid. 

\begin{figure*}
\centering
\includegraphics[width=15 cm]{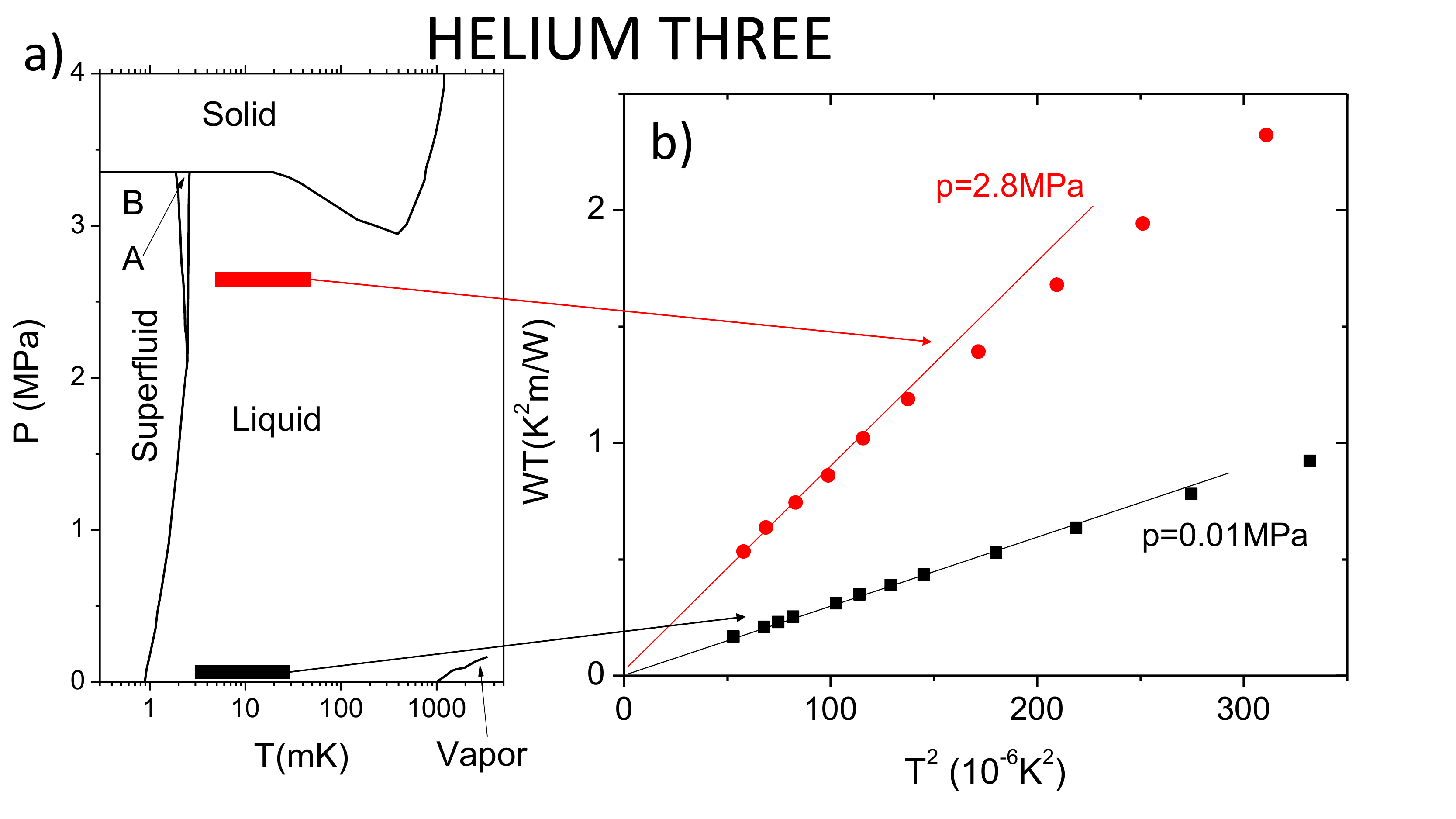}
\caption{\textbf{The original Fermi liquid:} a) Phase diagram of $^3$He \cite{Dobbs}. Black and red rectangles designate the temperature range of the two sets of data shown in the right panel. b) Thermal resistivity at two different pressures as a function of temperature according to the data reported by Greywall \cite{greywall1984}.}
\label{fig:3He}
\end{figure*}

The phase diagram of $^3$He is shown in Fig. \ref{fig:3He}a. The superfluid transition occurs below 3 mK. The normal liquid is  a degenerate liquid of fermions below 1 K. Under a pressure of $\approx$3 MPa (30 atmospheres), interaction between fermions becomes strong enough in order to solidify the liquid. Greywall measured the thermal conductivity $\kappa$ of the liquid down to $\sim 7$ mK at different pressures, ranging from vacuum up to the threshold of solidification. He found that, in the first approximation,  $\kappa$ is proportional to the inverse of temperature, as expected and quantified the product of thermal conductivity and temperature, $\kappa T$, at different pressures. 

Now, since $WT\equiv \frac{T}{\kappa}$, the expression $\kappa \propto T^{-1}$ is strictly equivalent to $WT \propto T^2$. Fig. \ref{fig:3He}b shows $WT$ extracted from Greywall's $\kappa (T)$ data at two different pressures (or molar volumes) as a function of $T^2$. One can see that at low temperature, thermal resistivity displays a T-square temperature dependence. At high pressure, when interactions are stronger, the slope (i.e. $B$ of Eq.\ref{WT}) is larger. Moreover, when the prefactor is larger, the temperature below which the T-square behavior is visible shifts to lower temperatures. These features do not surprise those familiar with the Kadowaki-Woods ratio in heavy-electron metals. For example, near the quantum critical point of CeCoIn$_5$ \cite{Bianchi2003,Paglione2003} and YRh$_2$Si$_2$ \cite{Gegenwart2002,Knebel2006}, the electronic specific heat and the T-square prefactor increase and the domain of validity of the T-square resistivity shrinks. 

\begin{figure*}
\centering
\includegraphics[width=15cm]{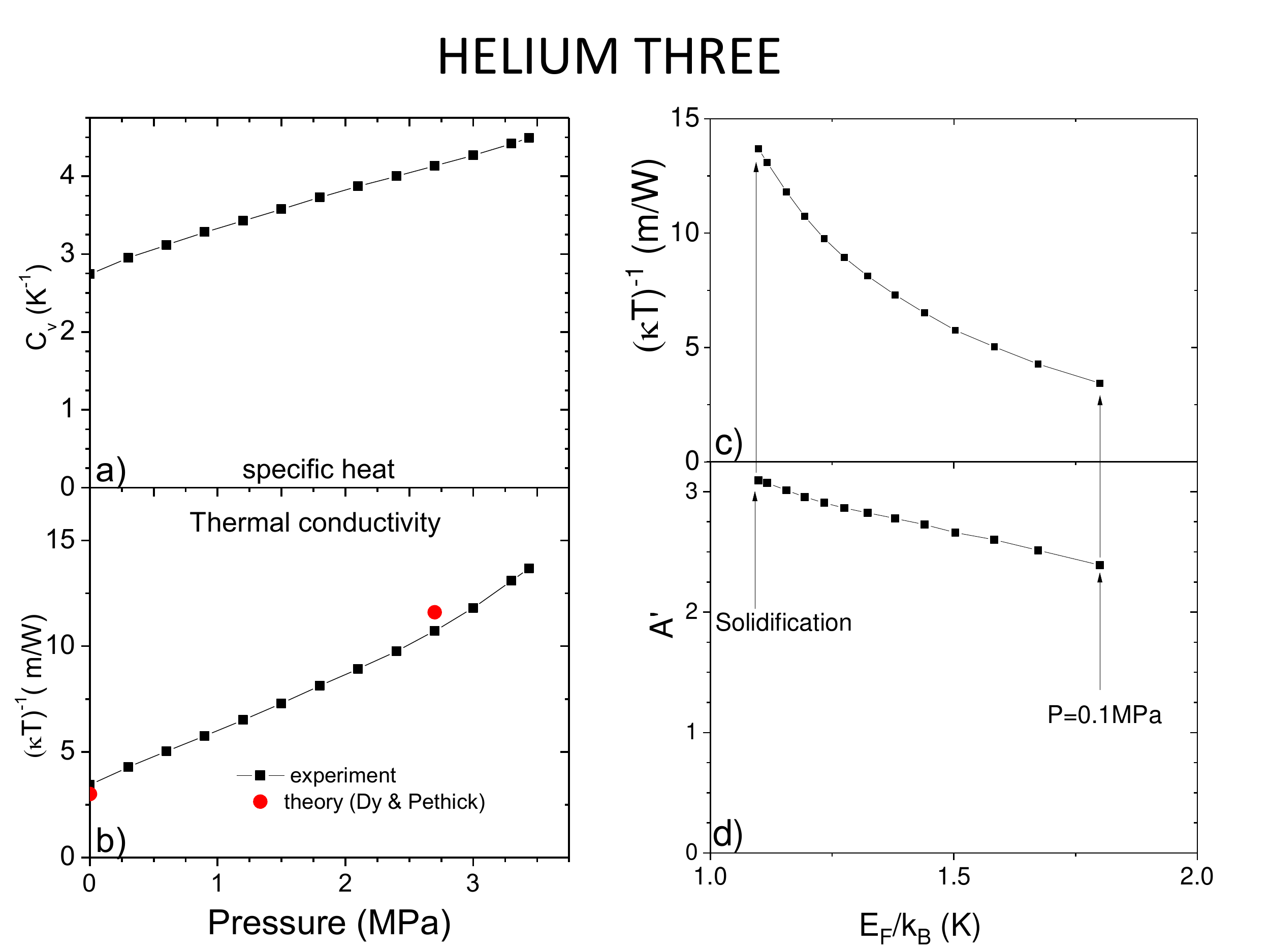}
\caption{\textbf{Tuning correlation with pressure-}  a) T-linear Specific heat as a function of pressure \cite{greywall1983}. b) $(\kappa T)^{-1}$ plotted against pressure, according to experiment (black solid circles) \cite{greywall1984} and theory (red solid circles) \cite{Dy1969}. c) The same quantity plotted as a function of the Fermi energy at each pressure.  d) The dimensionless parameter $A'$ (see text) as a function of the Fermi energy. In these panels, $(\kappa T)^{-1}$ and T-linear specific heat are values reported by Greywall who extrapolating the data to extract zero-temperature limit. } 
\label{fig:HekT}
\end{figure*}

With increasing pressure, interaction between $^3$He atoms intensify and above a threshold pressure, the liquid solidifies. This striking version of pressure-tuned correlation inspired P. W. Anderson to state that ``the Mott insulator is a form of quantum solid, and the melting transition in $^3$He is our best example of a Mott transition.'' \cite{Anderson1997}.

Fig. \ref{fig:HekT} shows how the specific heat and the fermion-fermion scattering (which is proportional to $(\kappa T)^{-1}$) evolve with pressure. One can see that both smoothly enhance as the solidification approaches. Red circles in Fig. \ref{fig:HekT} are values expected by theory \cite{Dy1969} which are close to the experimentally-measured data.

Calkoen and van Weert \cite{calkoen1986} observed that the magnitude of $(\kappa T)^{-1}$ \cite{greywall1984} allows to extract a Landau parameter, which is in excellent agreement with the one  extracted from the specific heat\cite{greywall1983}. They argued that at zero temperature limit, one has:
\begin{equation}
\label{kappaT2}
\kappa T|_0= \frac{5}{18\pi^3}\frac {p_F^3 v_F^2 } {A^2 }
\end{equation}

Here, $p_F$  and $v_F$ are Fermi momentum and Fermi velocity. The parameter $A$ (not to be mistaken with the prefactor of T-square resistivity) has the dimensions of the Planck constant and was identified as a combination of Landau Fermi-liquid parameters \cite{calkoen1986}.  Extracting the Fermi temperature from the specific heat per volume, one can monitor the evolution of $(\kappa T)^{-1}$ and $A'=\frac{A}{\hbar}$ with the Fermi temperature. As seen in Fig.\ref{fig:HekT}c-d, $(\kappa T)^{-1} $ steadily increases with the Fermi energy. This is concomitant with the evolution of dimensionless $A'$, intimately linked with the Landau parameter $F^a_0$ \cite{calkoen1986}.

Thus the amplitude and the pressure dependence of $\kappa T$ in $^3$He are well understood. The prefactor of T-square thermal resistivity, $B$, is the inverse of $\kappa T$ and therefore, Eq.\ref{kappaT2} can be rewritten as:

\begin{equation}
\label{kappaT3}
B\equiv (\kappa T)^{-1}=\frac{9\pi^3}{10}\frac {\hbar A'^{2}} {E_F^2 k_F }
\end{equation}

Here, $A$ is replaced by dimensionless $A'$  and the presence of the Fermi energy $E_F=\frac{1}{2}p_Fv_F$ is explicit. 

\begin{table}
\centering
\begin{tabular}{lcc}
\hline
Reference & $\tau^0_{\eta}T^2$ (ps.K$^2$)& $\tau^0_{\kappa}T^2$ (ps.K$^2$)\\
\hline
Greywall (1984) (p=0) \cite{greywall1984} & -- & 0.391 \\
Wheatley (1975)  (p=0)\cite{Wheatley1975} & 1.24 & 0.51 \\
Alvesalo \textit{et al.} (1975) \cite{alvesalo1975}&0.65& 0.23\\
Bertinat\textit{et al.} (1974) \cite{bertinat1974}&1.29& 0.524\\
Black \textit{et al.} (1971) \cite{black1971} & 1.58 & -- \\
\hline
\end{tabular}
\caption{The rate of fermion-fermion scattering extracted from measurements of viscosity and from measurements of thermal conductivity as reported by several authors. The time scale associated with momentum diffusion is 2-3 times longer than the scattering time associated with energy diffusion. Compare this with what is seen in metals (Fig. \ref{fig:WT}) and the dichotomy between vertical and horizontal events. }
\label{Tab2}
\end{table}
Several studies have been devoted to measuring the temperature dependence of viscosity$\eta$. They have confirmed that the expected  T-square decrease in the amplitude of viscosity  ($\eta \propto T^{-2}$). These studies allow quantifying the fermion-fermion scattering rate giving rise to the temperature-dependent viscosity ($\tau^0_{\eta}T^2$). Several authors have compared it then with the one extracted from thermal conductivity($\tau^0_{\kappa}T^2$). A summary of these numbers reported by various authors is given in table 2.  One can see that the scattering time for momentum diffusivity is 2-3 times longer than the scattering time for energy diffusivity. Like in the case of metals, and by a comparable factor,  thermal transport is punished more than momentum transport by fermion fermion scattering.

Normal liquid $^3$He demonstrates not only the persistence of T-square thermal resistivity in a single fermionic reservoir without Umklapp scattering, but also the scaling between its amplitude and the square of the Fermi energy. Let us now  check the relevance of these facts to metals.  

\section{A common thread}
\begin{figure*}
\centering
\includegraphics[width=15 cm]{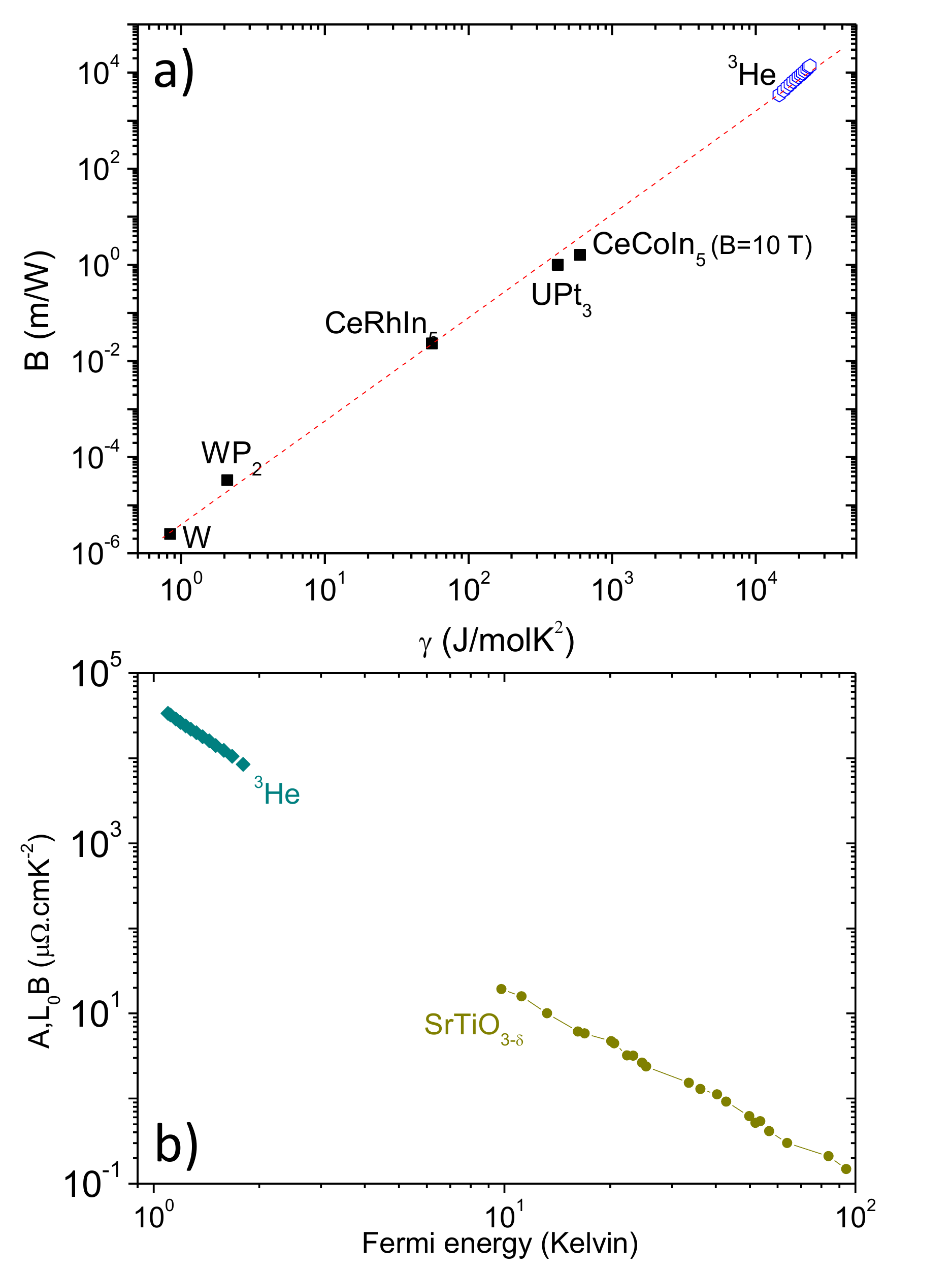}
\caption{\textbf{Continuity between $^3$He and metals:} a) The magnitude of T-square thermal resistivity,  $B$, in $^3$He and in metals where the T-square thermal resistivity has been measured plotted against the T-linear specific heat, $\gamma$. The dashed line, which is a guide to eye has a slope of 2.15. b) The T-square thermal resistivity prefactor multiplied by the Sommerfeld value ($L_0B$) in $^3$He and the T-square electrical resistivity prefactor ($A$) in lightly-doped single-band strontium titanate as a function of the Fermi temperature. }
\label{fig:Hemetal}
\end{figure*}

Fig. \ref{fig:Hemetal}a is a plot \`a la Kadowaki-Woods (or Rice), exclusively focused on thermal transport. Each data point represents a metallic solid in which the thermal resistivity prefactor $B$ has been quantified. It includes two weakly correlated (W \cite{wagner1971} and WP$_2$ \cite{jaoui2018}) and three strongly correlated  (UPt$_3$ \cite{lussier1994}, CeRhIn$_5$ \cite{paglione2005} and CeCoIn$_5$ \cite{paglione2006}) metals. Note that CeCoIn$_5$ is not a Fermi liquid at zero magnetic field. Its resistivity displays a T-square resistivity only in presence of a finite magnetic field restoring the Fermi-liquid behavior \cite{Paglione2003,Bianchi2003}. The plot compares the amplitude of $B$ in these metals and $B\equiv (\kappa T)^{-1}$ in $^3$He, discussed in the previous section, as a function of their $\gamma$. The data points scatter close to the dashed line which represents a $\gamma^{2.15}$ slope in the log-log scale. Note that  $^3$He data points evolves slightly faster, but reasonably close to the dashed line. 

Up to now, the amplitude of T-square thermal resistivity in metals has been analyzed starting from T-square electrical resistivity. It has been taken for granted that the latter cannot arise by normal (i.e not Umklapp), intra-pocket electron-electron scattering. Fig.\ref{fig:Hemetal}a which focuses on thermal transport, provides an alternative and a short-cut root. Normal momentum-conserving between fermionic quasi-particles generates T-square thermal resistivity in metals as it does in $^3$He. 

The idea that normal electron-electron scattering plays a role in generating $B$ has found support by the results of recent experiments on thermal transport in bulk antimony (Sb). In this semi-metal, electrons are quasi-ballistic with a mean-free-path  approaching (and scaling) with the sample size. Jaoui \textit{et al.}\cite{Jaoui2021} quantified the prefactors of the thermal ($B$) and the electric ($A$) T-square resistivities and found that the $BL_0/A$ ratio evolves with the change in residual resistivity (or sample size). The cleaner the sample, the larger was the $B/A$ ratio and the deviation from the WF law\cite{Jaoui2021}. This behavior finds a natural explanation in the hydrodynamic scenario of heat transport as the driver of the deviation from the WF law at the onset of ballistic regime. Indeed, according to a theoretical work by Principi and Vignale \cite{principi2015} the WF ratio between the thermal and the electric conductivity is reduced by a $1+\tau_0/\tau_{ee}$ factor, where $\tau_0$ is the momentum relaxing time and $\tau_{ee}$ is the electron-electron scattering time. In Sb, at the onset of ballistic regime  ($\sim 10 K$), these two time scales are comparable in amplitude. In cleaner samples, $\tau_0$ is larger and the deviation from the WF law is more pronounced. Thus attributing a role to normal electron-electron collisions in generating $B$ provides a satisfactory explanation to the evolution of the ratio of the two prefactors with sample size. In contrast, the traditional picture invoking horizontal and vertical events fails to explain why the deviation between thermal and electrical resistivity grows with sample size..

According to Principi and Vignale \cite{principi2015}, the WF law is recovered when $\tau_0 \ll \tau_{ee}$. In most metals, this condition is easily satisfied in a reasonable temperature range thanks to unavoidable disorder. In \textit{all} metals, the condition will be satisfied at sufficiently low temperature, because as $T\rightarrow 0$, $\tau_{ee}^{-1}$ will diverge and $\tau_{0}^{-1}$  will stay finite. Therefore, the T-square thermal resistivity should have an electrical counterpart of the amplitude set of $\sim L_0B$ and T-square electrical resistivity is expected even without Umklapp (or other momentum-relaxing) collision events between electrons. The combination of Normal electron-electron scattering and disorder will suffice for this.

In this picture, the (relative) success of the $A$ vs. $E_F^{-2}$ scaling across different families of metals is a consequence of the approximate validity of the WF law, which implies $A\approx \frac{\pi^2}{3}(\frac{k_B}{e})^2 B$. This, combined with Eq. \ref{A} and Eq.\ref{kappaT3}, yields the following expression for the phenomenological $\ell_{quad}$ :

\begin {equation}
\label{ell}
\ell_{quad}  \approx \frac{3\pi^5}{10}\frac{ A'^{2}} {k_F} 
\end{equation}

In the weakly-interacting metals considered by Rice \cite{Rice1968}  $\ell_{quad}\approx 1.6$ nm and in the strongly-correlated metals scrutinized by Kadowaki and Woods $\ell_{quad}\approx$ 40 nm \cite{lin2015,Wang2020}. Many other metals fall somewhere in between. According to Eq. \ref{ell}, the magnitude of $\ell_{quad}$ in $^3$He Mpa. 

Fig. \ref{fig:Hemetal}b) plots $L_0B$ in $^3He$ and $A$ in lightly-doped strontium titanate as a function of the Fermi energy. The two curves follow a similar trend.

\section{Concluding remarks}
Thus, when the thermal transport becomes the departing point, our puzzles find solutions. The link between the magnitude of $A$ and the Fermi energy becomes understandable. The persistence of T-square in dilute metals becomes unsurprising. 

The temperature dependence of resistivity in SrTiO$_{3-\delta}$ does not reduce to this issue. Well below the degeneracy temperature, the exponent of resistivity in this metal is close to two, but it  smoothly evolves with warming \cite{Lin2017,PhysRevX}. In order to explain  resistivity in a wide temperature range, one needs to invoke coupling between electrons and  transverse optical phonons as recently shown by two independent theoretical studies \cite{Kumar2021,Nazaryan2021} Nazaryan and Feigel'man \cite{Nazaryan2021}, in contrast to Kumar \textit{et al.}\cite{Kumar2021} , incorporated in their calculations the evolution of the effective mass with temperature found by experiment \cite{PhysRevX} and gave an account  of both resistivity and thermopower in a wide temperature range, starting from the minimum energy of phonons. Coupling with phonons cannot explain the persistence  of T-square resistivity below this temperature range, which unavoidably leads to electron-electron scattering. The present approach not only provides an explanation for this persistence, but also for its amplitude. 

Note that electronic thermal conductivity is orders of magnitude lower than the total thermal conductivity in SrTiO$_{3-\delta}$ \cite{martelli2018}. To check the validity of the WF law one needs a way to separate the two. This has been only done in the zero-temperature limit \cite{Lin2014}. Quantifying the electronic thermal resistivity and quantifying  $B$ in this system remains an experimental challenge. 

The case of graphene is beyond the scope of the present paper. In graphite, T-square resistivity emerges below 6 K\cite{Uher1977}. An investigation of resistivity in graphene, focused on the relevance of Bloch-Gr\"uneisen resistivity \cite{Efetov2010} did not explore such low temperatures. Theory expects the phase space of e-e scattering to follow a $T^2$ln$T$ temperature dependence \cite{Giuliani1982,Kaveh1984} in two dimensions. Recently, $T^2$ resistivity was observed in a Moir\'e superlattice of graphene on top of boron-nitride  \cite{Wallbank2019} and in twisted bilayer graphene \cite{Jaoui2021b}. Comparing the amplitude of T-square in these studies with what has been observed in three-dimensional solids emerges as a future task.

The distinction between e-ph and e-e scattering is not impermeable. There are cases where one suspects that scattering between electrons is accompanied by exchange of phonons \cite{Macdonald1980,Gurvitch1980,jaoui2022}. It has been suggested that phonon exchange amplifies the magnitude of $A$ in Al \cite{Macdonald1980} or in A15 compounds \cite{Kaveh1984}. It remains a task for theory to sort out the link between phonon exchange and the electron-electron cross-section. 

The  interplay between  disorder and electronic interaction is yet to be sorted out. The recovery of the WF law in the picture put forward by Principi and Vignale \cite{principi2015} requires a hierarchy between the two scattering time: $\tau_0 < \tau_{ee}$. Upon warming, $\tau_{ee}$ becomes steadily shorter and  eventually this inequality is no more satisfied. As seen in Fig. \ref{fig:examp}, the T-square behavior seen in several metals extends to temperatures above $AT^2 \sim \rho_0$. This implies a crucial missing component. 

The analogy between the thermal resistivity of metals and $^3$He is beyond dispute. But how relevant is momentum diffusivity to electrical resistivity of metals? The usual answer to this question is negative. The thermal current is accompanied by a gradient of thermal energy, but the drift velocity is identical along the charge current. In presence of ohmic response, the electron viscosity \cite{gurzhi1968} of a metallic Fermi liquid has seldom been considered relevant to diffusive electric transport. However, Landauer \cite{Landauer1975,Landauer1987} contested the assumption that current flows homogeneously in a macroscopic sample hosting defects. He postulated that, within a screening distance of each scattering center,  electric field and charge current are both inhomogeneous. 

Let us end by noting a number of studies, which may prove relevant to the subject matter of the present paper. Einzel and  Parpia have already examined the link between Drude conductivity in metals and the Poiseuille flow in normal liquid $^3$He in presence of aerogel \cite{Einzel1998}. Nazraov \textit{et al.} have shown that dynamical electron-electron interactions, known to govern the conductivity of nanoscale junctions \cite{Sai2005}, play a role in setting the zero-temperature resistivity of a bulk metal \cite{Nazarov2014}. Evidence for inhomogeneous charge flow is provided by experiments imaging the flow of current in a two-dimensional electron gas (2DEG) \cite{Topinka2001}, which have found that it follows branching strands driven by  ripples in the energy landscape. 

Thus, the viscosity of the electron liquid may be more accessible than previously thought, and the `hydraulic analogy' \cite{Herring1954} deeper than a mere metaphor.




\medskip
\textbf{Acknowledgements-} I am grateful to Beno\^it Fauqu\'e, Mikhail Fei'gelman, Alexandre Jaoui, Xiao Lin, Dmitrii Maslov, Vladimir Mineev, and Giovanni Vignale for their valuable comments. This work was supported by the Agence Nationale de la Recherche (ANR-18-CE92-0020-01; ANR-19-CE30-0014-04).
\medskip

%
\bibliography{Biblio}

\end{document}